# Label-free SERS Discrimination of Native Proline Hydroxylation at Single-molecule peptide by Deep Learning-assisted plasmonic nanopore

Yingqi Zhao[1,3],Kuo Zhan [1,3], Pei-Lin Xin[1,3], Yuge Liang[1,3], Enock Adjei Agyekum[4], Matti Putkonen[5], Shuai Li[4], Francesco De Angelis[6] and Jianan Huang [1,2,3,*]

[1] Research Unit of Health Sciences and Technology, Faculty of Medicine, University of Oulu, Aapistie 5 A, 90220 Oulu, Finland.

[2] Research Unit of Disease Networks, Faculty of Biochemistry and Molecular Medicine, University of Oulu, Aapistie 5 A, 90220 Oulu, Finland.

[3] Biocenter Oulu, University of Oulu, Aapistie 5 A, 90220 Oulu, Finland.

[4] The Biomimetics and Intelligent Systems (BISG) research unit, Faculty of Information Technology and Electronic Engineering, University of Oulu, Oulu, Finland

[5] Department of Chemistry, University of Helsinki, P.O.Box 55 (Chemicum, A.I. Virtasen aukio 1)

[6] Istituto Italiano di Tecnologia, Via Morego 30, 16163, Genoa, Italy.

*Email: jianan.huang@oulu.fi

**ABSTRACT:** Post-translational modifications (PTMs) play essential roles in regulating protein structure, function, and cellular signaling. Hydroxylation of hypoxia-inducible factor-1α (HIF-1α) is a biologically important PTM involved in oxygen-dependent regulation and cancer-related pathways; however, peptide-level discrimination of hydroxylation at the single-molecule level remains difficult. Here, we report a particle-in-pore single-molecule surface-enhanced Raman spectroscopy (SERS) platform combined with peak occurrence frequency (POF) analysis and a one-dimensional convolutional neural network (1D-CNN) for discriminating hydroxylated and non-hydroxylated HIF-1α peptide fragments. Three peptide pairs containing the Pro-564 hydroxylation site, with lengths of 7, 9, and 15 amino acids (AAs), were investigated. POF analysis revealed reproducible hydroxylation-dependent spectral changes in the 7AA and 9AA peptide pairs, which were attributed to changes in adsorption conformation and surface interactions. CNN-based classification achieved post-evaluation accuracies of 72.98%, 78.55%, and 89.74% for the 7AA, 9AA, and 15AA peptide pairs, respectively, with AUC values above 0.80 for all the pairs, indicating a reliable discrimination. Gradient-weighted feature visualization further showed that CNN-sensitive regions overlapped with recurrent POF features, supporting the chemical relevance of the learned classification patterns. Notably, for the 15AA peptide pair, the enhanced citrate-associated band suggests that hydroxylation can substantially alter peptide–gold nanoparticle adsorption behaviour. This adsorption-mediated effect may amplify hydroxylation-induced spectral differences and contribute to the improved discrimination accuracy despite the increased structural complexity. These results demonstrate that the particle-in-pore sensor, assisted by deep learning, can capture hydroxylation-induced spectral and adsorption changes in HIF-1α peptide fragments, providing a promising strategy for ultrasensitive analysis of weak PTM signatures in peptides.

Post-translational modifications (PTMs) are essential regulators of cellular function, as they can profoundly influence protein structure and conformational dynamics. Among these modifications, prolyl 4-hydroxylation involves the enzymatic addition of a hydroxyl group to proline residues and plays a central role in the regulation of hypoxia-inducible factor (HIF).[1, 2] Through an oxygen-dependent mechanism, HIF controls the expression of hundreds of genes implicated in cancer progression. This landmark discovery was recognized with the 2019 Nobel Prize in Physiology or Medicine. As a key modulator of tumor biology, HIF hydroxylation has emerged as a promising biomarker for evaluating tumor behavior and therapeutic response, as well as a potential target for intervention.[3] Despite this potential, existing detection approaches, including mass spectrometry- and antibody-based methods, remain limited in their ability to sensitively identify low-abundance PTMs. Therefore, single-molecule analysis of trace PTMs could provide significant advantages for early disease diagnosis and the detection of low-abundance protein species. [4]

Mass spectrometry is currently the dominant technology for PTM analysis. However, its sensitivity is constrained by the need for $10^6$–$10^8$ molecular copies, and PTM identification often suffers from false interpretation due to altered combinatorial rules. Nanopore resistive pulse sensing has emerged as a promising single-molecule detection method, distinguishing biomolecules based on current changes as they pass through a nanopore.[5] Using this technique, single-molecule PTM detection has been demonstrated for phosphorylation,[6] acetylation,[7] propionylation,[8] glycosylation,[9] nitration and oxidation.[10] More recently, the discrimination of peptide proline hydroxylation was demonstrated on a specifically engineered biological nanopore that interacts with the hydroxylation site and induces a change in the ionic current signature.[11] However, nanopore readouts are inherently indirect, being encoded in analyte-dependent blockade amplitudes, dwell times, and interaction dynamics with the inner pore environment.

Surface-enhanced Raman spectroscopy (SERS) offers direct access to vibrational fingerprints associated with, molecular structure, local chemical bonding and microenvironmental changes.[12-14] Most previous SERS studies, however, have centered on PTMs with relatively strong Raman cross-sections or easily distinguishable spectral features, including phosphorylation, [15, 16] and nitration, [17] and oxidation.[18] By comparison, hydroxylation and other PTMs with weak Raman signatures have received far less attention, largely because their spectral contributions are difficult to resolve. [19, 20] Moreover, signals arising from these subtle modifications can be masked by overlapping bands from the molecular backbone.[2] Meanwhile, variations in peptide adsorption configurations and molecular orientations on metallic surfaces, together with the heterogeneous distribution of localized electromagnetic fields at SERS hotspots, can further amplify spectral fluctuations. As a result, conventional discrimination strategies based on peak position comparison and peak assignment often suffer from limited reproducibility and difficulties in spectral interpretation. These issues have been recognized in previous reviews as the major bottlenecks in SERS-based analysis of protein PTMs, including weak signal intensity, spectral overlap, substrate and hotspot heterogeneity, and the inherent difficulty of structural interpretation in complex biomolecular systems.[2, 21]

In contrast to the above methods, recent plasmonic nanopore technologies demonstrated single-molecule SERS (SM-SERS) discrimination of 20 amino acids.[22] For example, by constructing a plasmonic nanogap between a gold nanoparticle and nanopore side wall in a flow-through setup, the particle-in-pore system enabled the shaping of a plasmonic hot spot with single-molecule sensitivity and specially confined hotspot to resolve amino acid (AA) bases and minimise backbone signal coverage.[22-24] Nevertheless, because the hot spot of this strategy can cover segment of 1 to 3 AAs of the peptide, the major SERS signals come from the non-PTM sites of the peptide and would challenge identification of the peptide PTMs. To solve this problem, machine learning (ML) models have been used for SERS spectral analysis, enabling automated recognition of complex spectral patterns that are difficult to interpret by conventional peak assignment.[25, 26] Recently, machine-learning-assisted identification of the short peptides vasopressin and oxytocin, which differ by only two amino acids, was reported in plasmonic nanopore systems. Using a Random Forest classifier, classification accuracies ranging from 52% to 80% were achieved.[27] Considering the small hydroxyl group, the detection of hydroxylation in peptides remains considerably more challenging because of the greater structural complexity of peptides, their heterogeneous adsorption behaviour, and analyte-dependent conformational variations on the sensing surface.

In this study, we demonstrate discrimination of proline hydroxylation in native HIF-1α peptide fragments by combining particle-in-pore single-molecule SERS with peak occurrence frequency (POF) analysis and a one-dimensional convolutional neural network (1D-CNN). The 1D-CNN model has been demonstrated to discriminate the proline from its hydroxylation. [28] Three pairs of HIF-1α peptide fragments containing the Pro-564 hydroxylation site, with lengths of 7AA, 9AA, and 15AA, were investigated to evaluate the influence of peptide length and structural complexity on discrimination performance. POF analysis revealed reproducible hydroxylation-dependent spectral changes, suggesting that proline hydroxylation alters peptide adsorption conformation and surface interactions on gold nanoparticle(AuNPs). CNN-based discrimination achieved post-evaluation single-spectrum classification accuracies of 72.98%, 78.55%, and 89.74% for the 7AA, 9AA, and 15AA peptide pairs, respectively, with AUC values above 0.80 for the 7AA and 9AA peptides and above 0.96 for the 15AA peptide. Gradient-weighted feature visualization further showed that CNN-sensitive spectral regions overlapped with recurrent POF features, supporting the chemical relevance of the learned classification patterns. Notably, for the 15AA peptide pair, hydroxylation induced a more pronounced change in peptide–AuNP adsorption behavior, reflected in the displacement of the native citrate capping layer on the AuNP surface. Collectively, these results demonstrate that particle-in-pore single-molecule SERS, assisted by statistical spectral reconstruction and deep learning, can capture subtle hydroxylation-induced spectral and adsorption changes, providing a promising framework for ultrasensitive analysis of weak PTM signatures in peptide systems.

## Results and Discussion

### Single-Molecule SERS Detection of HIF-1α Fragment Peptides

Figure 1(a) shows a schematic of the particle-in-pore system, which consists of a gold nanohole array with 250 nm diameter pores fabricated in a 100 nm thick gold film, which is supported by a free-standing SiN membrane on a silicon frame. The detailed fabrication procedure is described in the Method. A representative SEM image of a nanopore is shown in Figure 1(b). The SiN membrane chip is integrated into a microfluidic device with top and bottom reservoirs and corresponding electrodes. Before measurement, the analyte peptides are physically adsorbed onto 50 nm gold nanoparticles and introduced into the bottom reservoir of the microfluidic device. A transmembrane potential was applied by the top and bottom electrodes to drive the negatively charged AuNPs through the nanopore. Meanwhile, a 785 nm laser incident from above exerts an optical force that pushes the nanoparticle toward the nanopore sidewall. Under the combined action of electrophoresis, electroosmosis, and optical force, the nanoparticle can be trapped near the sidewall for an extended period. This process positions the analyte within the plasmonic hotspot formed between the AuNP and the nanopore sidewall, where the highly localized electromagnetic field excites the molecule and produces the enhanced Raman signal, as illustrated in the scheme in Figure 1(c). [23] The pink small dots in Figure 1(c) indicate the citrate capped on the AuNP which served as surfactant and reducing agent in the AuNP fabrication.

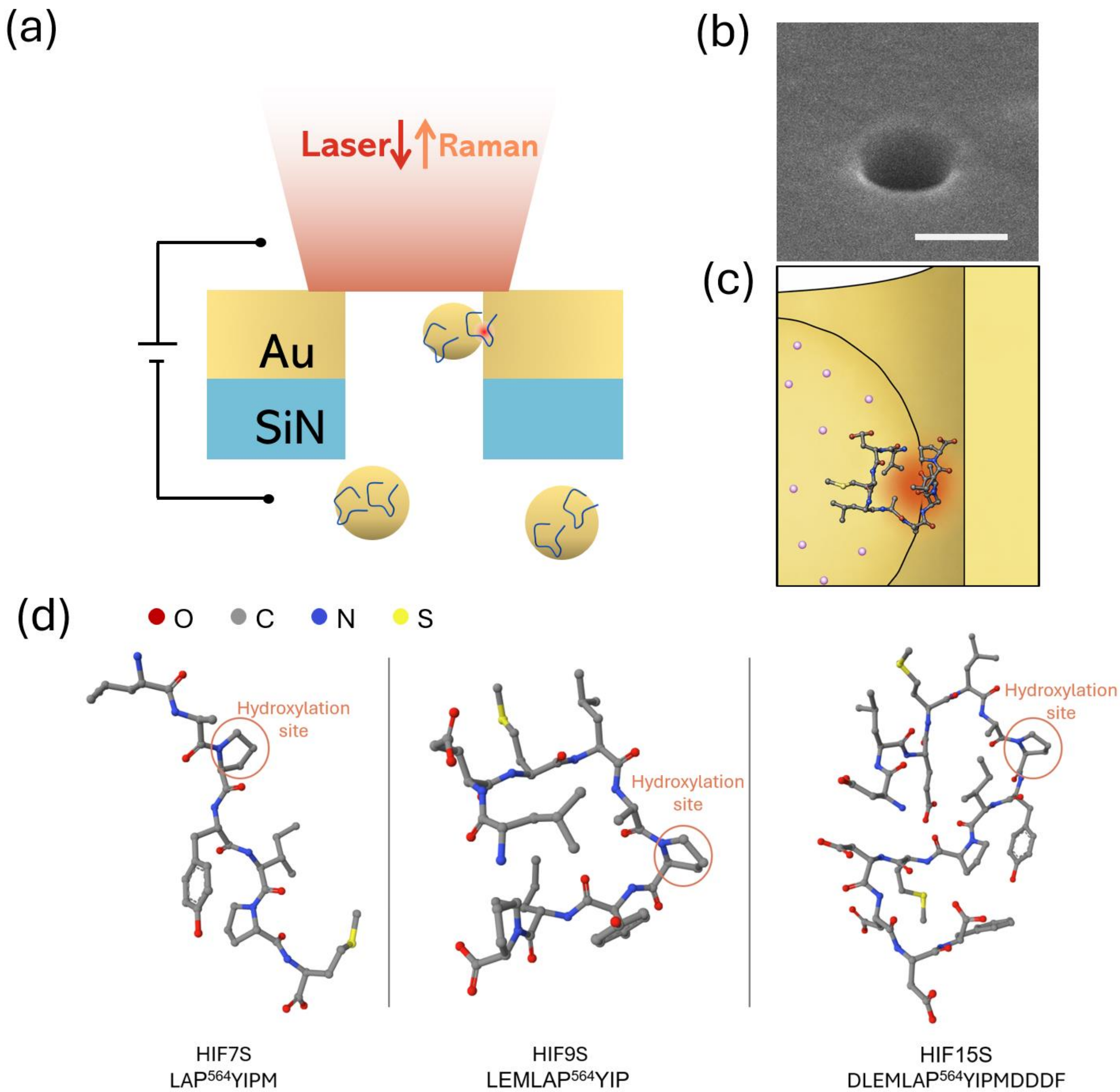

***Figure 1****. (a) Schematic illustration of the particle-in-pore sensor that allows single AuNPs loaded with peptides to be trapped in a gold nanopore with plasmonic resonance upon laser excitation at 785 nm. (b)SEM image of the nanopore. The scale bar is 250nm. (c) Schematic illustration of a peptide in the hotspot formed between a trapped AuNP and nanopore sidewall, the pink small dots on the AuNP indicates the citrate. (d)The molecular structure of peptides HIF7S, HIF9S and HIF15S. The Pro sites on which hydroxylation happen are indicated by orange circles. The oxygen, carbon, nitrogen, and sulfur atoms are indicated by red, gray, blue and yellow respectively.*

Three pairs of HIF-1α fragment peptides containing the Pro-564 residue, with lengths of 7AA, 9AA, and 15AA, were tested for discrimination of proline hydroxylation. Their names, sequences, and positions within the original HIF-1α protein are listed in Table 1. The HIF-1α sequence information was obtained from the UniProt Knowledgebase (UniProtKB), using the human HIF-1α entry Q16665. The non-hydroxylated 7AA, 9AA, and 15AA peptides were named HIF7S, HIF9S, and HIF15S, respectively, while the corresponding peptides containing hydroxylated Pro-564 were named as HIF7SPTM, HIF9SPTM, and HIF15SPTM. The AlphaFold-predicted[29] structures of HIF7S, HIF9S, and HIF15S are shown in Figure 1(c). Pro-564, the hydroxylation site, is indicated by an orange circle. Detailed information on the AlphaFold3 modelling is provided in Supplementary Note 1 and Table S1, in the Supporting Information. HIF7S displayed an almost linear predicted structure, whereas HIF9S began to adopt a U-shaped conformation, likely due to the presence of two proline residues and the increased peptide length. With further extension of the peptide chain, HIF15S exhibited a more coiled three-dimensional conformation, with Pro-564 oriented outward. Because these molecules are short peptides, the AlphaFold models were interpreted as putative conformational models rather than definitively determined structures. Nevertheless, the predicted conformations suggest that longer peptides may interact with AuNPs through preferentially exposed residues or specific binding regions, rather than adsorbing homogeneously along the entire peptide sequence.

***Table 1****. Peptide name, sequence and their position in HIF-1α*

| Peptide | Peptide Sequence | Position in HIF-1α |
|---|---|---|
| HIF7S | LAPYIPM | Residue 562-568 |
| HIF7SPTM | LA(Hyp)YIPM | |
| HIF9S | LEMLAPYIP | Residue 559-567 |
| HIF9SPTM | LEMLA(Hyp)YIP | |
| HIF15S | DLEMLAPYIPMDDDF | Residue 558-572 |
| HIF15SPTM | DLEMLA(Hyp)YIPMDDDF | |

A peptide monolayer was formed on AuNPs by incubating the nanoparticles with peptide-containing 5% PBS buffer at pH 7.4 for 48h. After incubation, the particle size and zeta potential were measured by Dynamic Light Scattering (DLS) to confirm that no serious aggregation had happened. The values of particle size and zeta potential are listed in the Supporting Information Table S2. The peptide concentration required to achieve monolayer coverage on the AuNP surface was estimated from the empirical maximum solvent-accessible surface areas of the peptides.[30] and the total nanoparticle surface area, as detailed in Supplementary Note 3. In this calculation, the peptide conformation and adsorption sites were not considered; instead, all residues were assumed to be in contact with the gold surface. Based on this assumption, the projected area of each peptide and the number of peptides required to form a monolayer on each AuNP were calculated. The purchased AuNPs were initially stabilized by a citrate layer, which acted as a surfactant. Upon incubation in peptide solutions, the citrate was expected to be replaced by peptides considering that the binding affinity of the citrate is much lower than that of amino acid residues.[31],[32] However, complete citrate replacement was not always observed in our experiments. This may be attributed to a reduced effective peptide occupied area when adsorption occurs through specific binding residues or local binding regions, rather than through uniform contact of the entire peptide with the AuNP surface. Alternatively, peptide conformation and hydroxylation-dependent changes in adsorption behaviour may also influence surface coverage, as discussed in the following sections.

**POF Analysis Reveals Hydroxylation-Dependent Spectral and Adsorption Changes**

The particle-in-pore sensor enables single-molecule sensitivity, but spectral interpretation is complicated by fluctuations and blinking typical of single-molecule SERS. During each trapping event, only part of the molecule may be located within the hotspot between the particle and the nanopore side wall, resulting in orientation and position-dependent excitation of selected vibrational modes. Aggregating large datasets allows a representative spectra profile to be reconstructed. Peak intensities are highly variable due to the relative position of the analyte molecule and the hot spot; they are less reliable for structural comparison. Statistical analysis of peak positions and occurrence frequencies across many events can reveal characteristic spectral features and structural changes.[33-35]

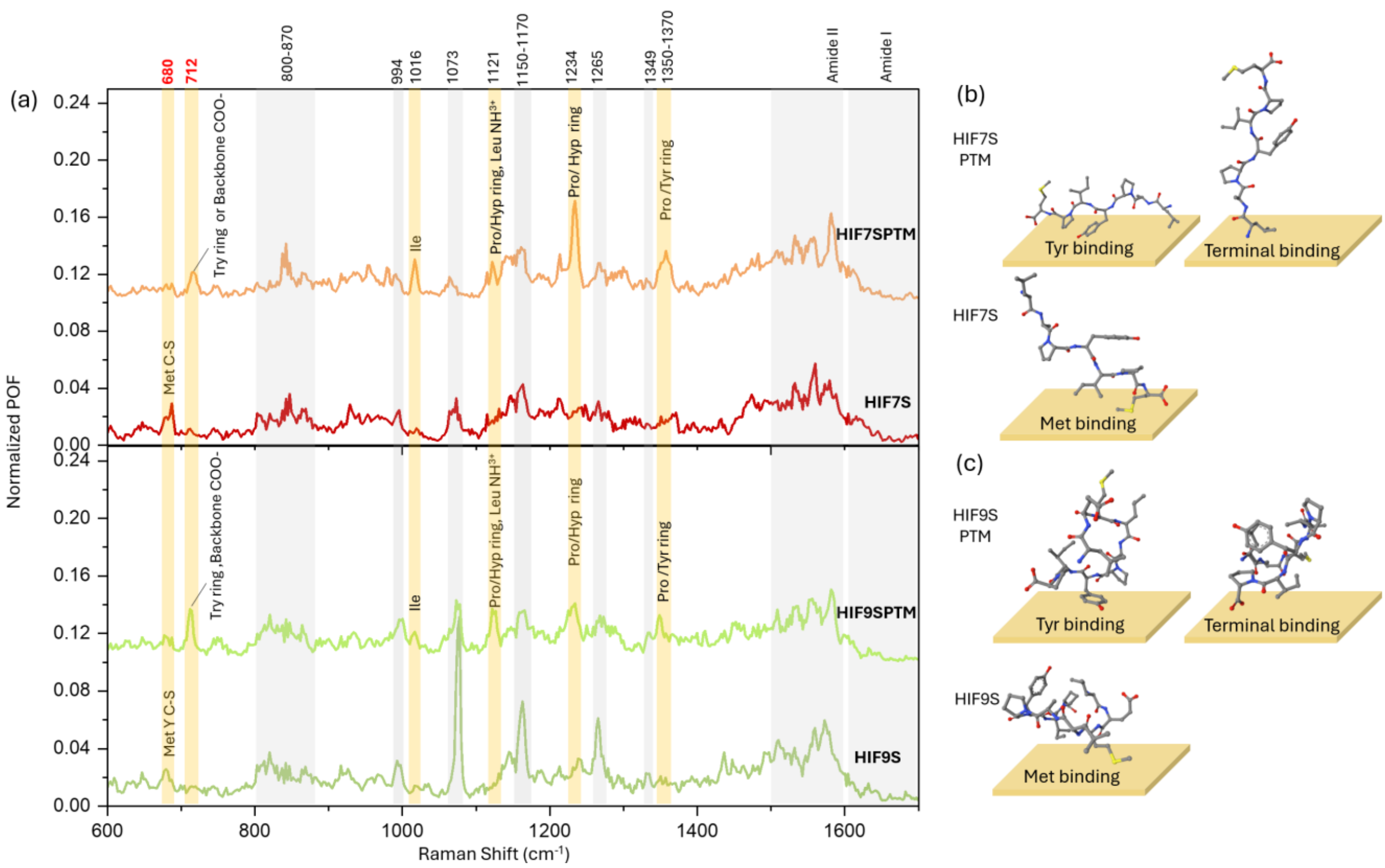

***Figure 2.*** *(a)Distribution of the normalized POF of HIF7S, HIF7SPTM, HIF9S, and HIF9SPTM, shown in red, orange, dark green, and light green, respectively. The peaks discussed in the text are highlighted by yellow and gray bands and labelled with their corresponding Raman shifts. Yellow-shaded bands indicate the region with similar spectra changes for 7AA and 9AA peptide pairs and can be tentatively assigned, whereas gray-shaded bands indicate other spectral regions where mentioned in the discussion. The offset of the HIF7SPTM and HIF9SPTM POF curve is 0.1. Proposed adsorption orientations of (b) HIF7S and HIF7SPTM and (c) HIF9S and HIF9SPTM on the gold surface are illustrated in the figures. The yellow plates indicate the gold surface.*

POF spectra provide an efficient approach for obtaining an overview of large single-molecule SERS datasets.[35] We quantified the peak occurrence frequency at each Raman shift and plotted the POF distribution spectra of HIF7S, HIF7S-PTM, HIF9S, and HIF9S-PTM, as shown in Figure 2(a). The detailed calculation of the POF distribution is described in the Methods. Briefly, the raw spectra were pre-processed by baseline correction, cosmic ray removal, normalization, and smoothing. The number of peak occurrences at each Raman shift was then calculated and divided by the total number of effective spectra. More than 6800 effective spectra were collected for each dataset, and the number of effective spectra for each dataset is listed in Table S3 in the Supporting Information. Since peaks in the POF profile represent the collective distribution of peak positions across the entire dataset, extracted from individual spectra, slight peak shifts of several wavenumbers may occur. Therefore, peaks appearing within a small Raman-shift window can still be considered as the indicators of the same vibrational mode. In this work, a tolerance of approximately ±10 $cm^{-1}$ was used for tentative peak assignment.

When comparing the POF spectra of peptides with and without hydroxylation, similar spectral changes were observed for both the 7AA and 9AA peptide pairs, as highlighted by the yellow bands in Figure 2(a). For the non-hydroxylated peptides, a peak appears at approximately 680 $cm^{-1}$, whereas this band becomes much less pronounced after hydroxylation. In contrast, a prominent peak near 712 $cm^{-1}$ is observed for the hydroxylated peptides. Other clear and reproducible PTM-induced spectral changes are observed at approximately 1016, 1121, 1234, and 1357 $cm^{-1}$, where pronounced peaks appear in the hydroxylated peptides but are absent or much weaker in the corresponding non-hydroxylated peptides. The fact that these spectral changes are repeated in both the 7AA and 9AA datasets suggests that proline hydroxylation alters the peptide adsorption orientation on the AuNP surface.

The peak at 680 $cm^{-1}$ is assigned to the Met ν(C–S) vibration.[36, 37] Considering the strong affinity of the sulfur atom in methionine for gold surfaces, the non-hydroxylated peptides are likely to adsorb onto the AuNP surface through the Met residue, as illustrated in the lower panels of Figure 2(b) and (c). This adsorption configuration explains the high occurrence frequency of the Met ν(C–S) mode.

After hydroxylation, adsorption through Met is no longer dominant. Instead, the emergence of the peak near 712 $cm^{-1}$ suggests the involvement of new adsorption sites. Here, we propose two possible explanations for this PTM-induced change in binding behaviour, both of which are consistent with the observed spectral changes. First, the peak at 712 $cm^{-1}$ may originate from Tyr ring deformation.[38, 39] Among the residues present in the peptides studied here, the aromatic ring of tyrosine can exhibit enhanced affinity toward gold surfaces through π–metal interactions, as supported by spectroscopic shifts, adsorption-induced conformational changes, and strong surface-enhanced Raman signals.[40] If Tyr becomes the dominant adsorption site, vibrational modes from its neighbouring amino acid residues would be expected to become more pronounced. The peak at 1016 $cm^{-1}$ is assigned to Ile ρ(Cδ-asym)[37, 41, 42] and Ile is located near Tyr in the peptide sequence. Hyp is another neighbouring residue of Tyr; therefore, the peaks at 1121 and 1234 $cm^{-1}$ may be attributed to Hyp τ($CH_2$) and Hyp δ(ring) vibrations, respectively.[19] The broad band at 1350–1370 $cm^{-1}$, especially in HIF7S-PTM, may contain contributions from Pro, Tyr, Glu, Leu, and Met vibrations.[37] However, ring-related vibrational modes from Tyr and Hyp tend to generate stronger single-molecule SERS signals, increasing the probability of detectable peaks and thus increasing the POF in this region.

A second possible explanation is that the peptide adsorbs through terminal groups. For a peptide chain, the $COO^-$ and $NH_3^+$ groups at the backbone termini may also serve as anchoring sites on the gold surface, resulting in an end-on adsorption configuration.[43] In this case, the peak near 712 $cm^{-1}$ may originate from $COO^-$ deformation, while the peak at approximately 1124 $cm^{-1}$ may be assigned to Leu $NH_3^+$[43], as these groups are located near the peptide termini. In addition, the peak near 1243 $cm^{-1}$ may arise from Pro ω($CH_2$) [44], since Pro is located at the C-terminus of HIF9SPTM and near the C-terminal Met in HIF7SPTM. In this terminal-binding configuration, illustrated in the upper panels of Figure 2(b) and (c), the peptide is anchored through the chain end, making residues at or near the terminus more likely to contribute detectable SERS peaks.

Due to the complexity of peptide SERS spectra and the multiple possible peak assignments, we don't have strong evidence to preclude either of the above-mentioned possibilities based only on the POF distribution. Nevertheless, the clear and reproducible changes in the POF spectra provide useful spectral fingerprints for discriminating hydroxylated and non-hydroxylated peptides. In summary, proline hydroxylation appears to suppress Met-dominated adsorption and promote alternative adsorption configurations involving Tyr and/or the peptide termini.

Several additional spectral regions contain rich structural information and are therefore worth discussing, as indicated by the gray bands in Figure 2(a). Citrate shows characteristic peaks near 1073 $cm^{-1}$, which are assigned to Cit ν(CO).[45, 46] In HIF9S and HIF9S-PTM, the 1073 $cm^{-1}$ citrate-related peak remains relatively strong, indicating that citrate displacement by the peptide is less sufficient. A group of peaks between 820 and 870 $cm^{-1}$ may arise from Tyr, Pro, and Hyp ring-related vibrations. Similarly, the amide II region from 1500 to 1600 $cm^{-1}$ exhibits high peak occurrence and clear spectral variation. Manual peak assignment in this region is challenging because of the substantial overlap among vibrational modes; however, these spectral features remain informative. The Amide I region shows lower POF, in agreement with the previous report that for short peptides, the existence of bulky side chains, such as Tyr and Pro, will reduce the probability of observing the Amide I band.[47] In addition, several common peak regions were observed in both the hydroxylated and non-hydroxylated peptide spectra, including those near 994, 1150–1170, and 1265 $cm^{-1}$. These bands may arise from overlapping vibrational modes of multiple residues, and the possible assignments are listed in Table 2. Although these peaks are not unique to either peptide form, their relative occurrence frequencies and subtle spectral variations may still provide useful information for classification. Since the full spectra were used as inputs to CNN training, these regions may also contribute to the deep learning model's learned classification features, as discussed later. The POF-based peak assignments for HIF7SPTM, HIF7S, HIF9SPTM, and HIF9S are summarized in Table 2.

***Table 2**. Tentative peak assignment for HIF7SPTM, HIF7S, HIF9SPTM and HIF9S.**

| HIF7SPTM Band ($cm^{-1}$) | HIF7S Band ($cm^{-1}$) | HIF9SPTM Band ($cm^{-1}$) | HIF9S Band ($cm^{-1}$) | Tentative Assignment |
|---|---|---|---|---|
| | 680 | | 682 | Met ν(C-S)[36, 37] |
| 717 | | 712 | | Tyr ring[38, 39]; Backbone COO- def [43] |
| 800-870 | 800-870 | 800-870 | 800-870 | Pro ρ(CH2)[44]; Tyr ring breath, fermi doublet[48] |
| 994 | 995 | 997 | 993 | Pro ρ (CH2) [20, 44]; Tyr ω(CC), ω (CH) [49] |
| 1016 | | 1016 | | Ile ρ(C*δ*-asym) [37, 41, 42] |
| 1066 | 1071 | 1072 | 1073 | Cit ν(CO) [45, 46] |
| 1121 | | 1124 | | Hyp t(CH2)[19]; Leu $NH_3^+$ [43] |
| 1150-1170 | 1150-1170 | 1150-1170 | 1150-1170 | Cit[45], Glu, Leu, Met[37] |
| 1234 | | 1231 | | Pro ω(CH2)[44]; Hyp δ ring[19]; Tyr OH def [48] |
| 1265 w | 1265w | 1268 | 1266 | Hyp δ (CH2 OH)[19]; Met, τ(CH2) [37]; Glu[37] |
| 1349 | 1349 | 1335 | 1337 | Pro ω(CH2)[20]; Hyp δ(CH)[19]; Tyr ν(CN), δ(CH)[48]; Ile ω(Cβ)[42] |
| 1350-1370w | 1350-1370w | 1350-1370 | 1350-1370 | Pro, Tyr, Glu, Leu, Met [37] |

*For the spectral region1150-1170cm-1 and 1350-1370 with overlapping signals from multiple residues and multiple vibration modes from the same residues, only the residue name is listed instead of assigning a specific vibration band.

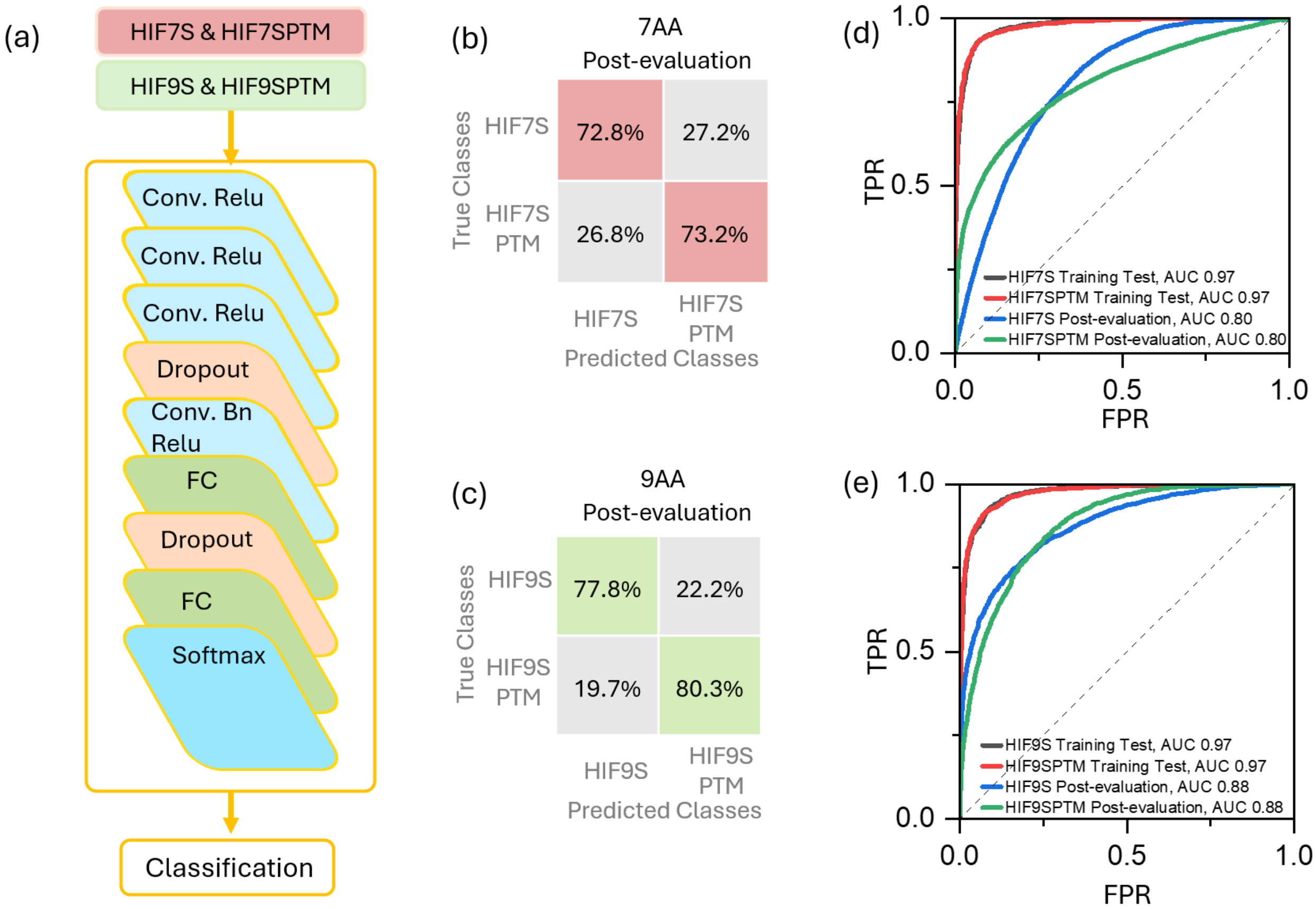

***Figure 3****. (a) Schematic representation of the CNN model architecture and data classification framework. (b) Post-evaluation confusion matrix for HIF7S and HIF7SPTM, (c) Post-evaluation confusion matrix of HIF9S and HIF9SPTM. (d) ROC curves and AUC values for HIF7S and HIF7SPTM testing and post-evaluation datasets. (e) ROC curves and AUC values for HIF9S and HIF9SPTM testing and post-evaluation datasets.*

**CNN-Based Discrimination.** Deep learning-based models have been widely applied to spectroscopic data analysis due to their ability to extract features and perform robust pattern recognition.[50, 51] We employed a 1D-CNN model, implemented in MATLAB, for the classification and post-evaluation of hydroxylated and non-hydroxylated peptides. For each dataset, more than 6800 effective spectra, each containing 981 spectral features, were used for CNN-based classification training. The spectra were divided into training and testing sets, with 80% used for training and 20% used for testing. The workflow included three main steps: dataset preparation, model training, and performance evaluation. In addition, more than 2200 independent spectra were used for post-evaluation for each peptide system. Due to variations in single-molecule trapping events, the number of effective spectra extracted from different measurements varied among datasets. The detailed numbers of effective spectra used for training, testing, and post-evaluation are provided in Table S3 in the Supporting Information.

The architecture of the CNN model is shown in Figure 3(a). The corresponding post-evaluation confusion matrices for the 7AA and 9AA peptides are shown in Figure 3(b) and (c). For the discrimination of the 7AA peptide, the model demonstrated single-spectrum accuracy higher than 72.98% both in the training test and post-evaluation. For the 9AA peptide, accuracy is higher than 78.55% in the training test and post-evaluation set. Receiver operating characteristic (ROC) curves are shown in Figure 3(c). The area under the curve (AUC) values for the 7AA and 9AA peptides are 0.80 and 0.88, respectively. The combination of post-evaluation accuracy above 72.98% and AUC values above 0.80 indicates meaningful discrimination capability and supports the reliability of the CNN-based classification. This result is particularly encouraging because the classification was performed under single-molecule SERS detection conditions, where spectra are intrinsically heterogeneous due to molecule-to-molecule variations in adsorption orientation, local electromagnetic enhancement, hotspot position, and conformational state. Moreover, hydroxylation represents a subtle chemical modification, and its spectral contribution may be weak relative to the signals from the rest of the peptide chain.

It should also be noted that the reported accuracy was obtained from single-spectrum classification. However, in the particle-in-pore platform, single-molecule SERS data are naturally collected as time-series spectral batches during nanoparticle trapping events, and model training and evaluation rely on a large number of spectra. Therefore, bag-level or ensemble-based decision strategies, which integrate predictions from multiple spectra collected from the same peptide sample or trapping event, may provide an even more suitable framework, which reflects the batch-based nature of the data for future analysis. Overall, the observed discrimination performance suggests that the CNN model was able to extract hydroxylation-associated spectral patterns despite substantial intra-class variability and inter-class spectral overlap.

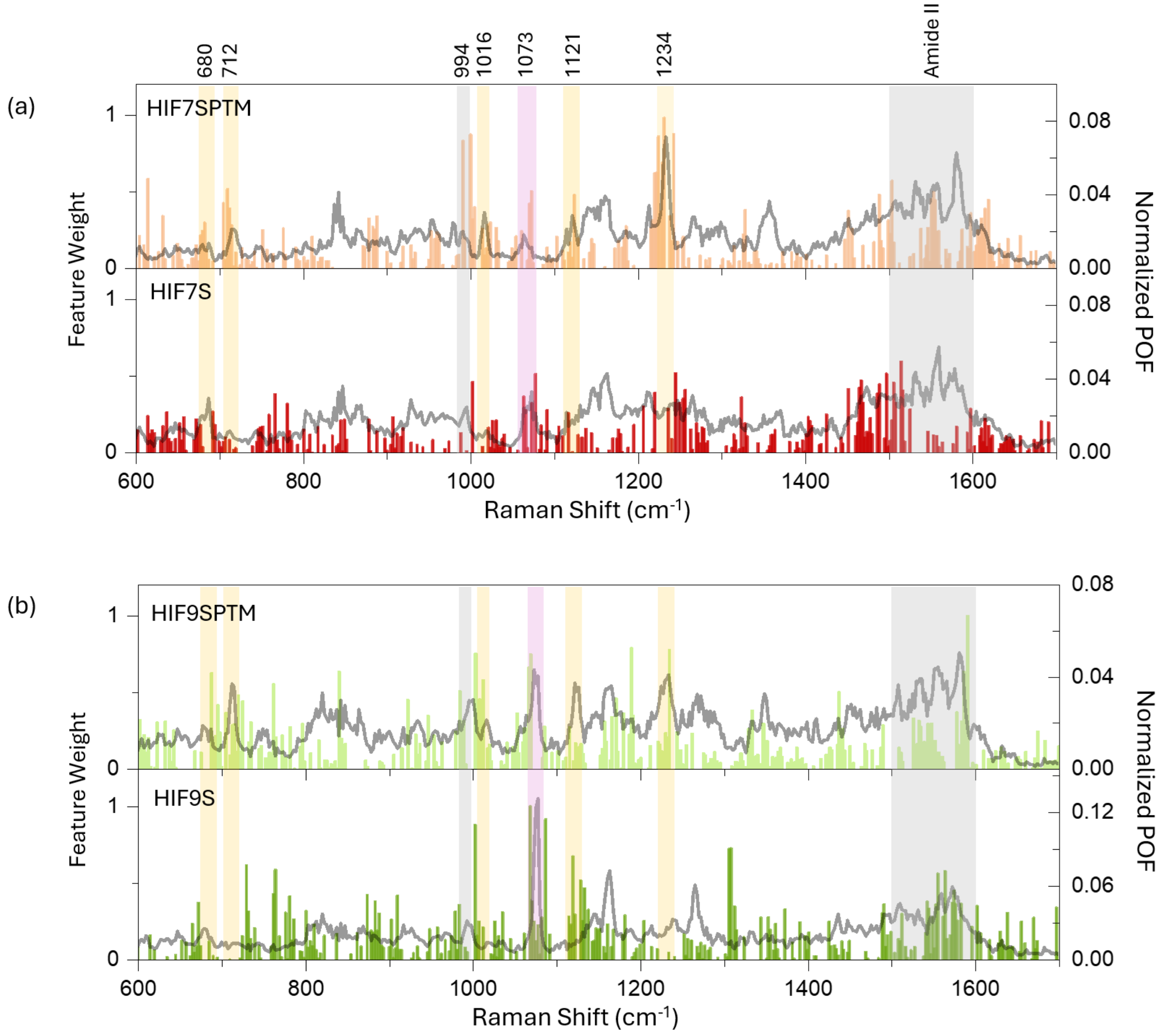

***Figure 4**. The feature weight extracted from the 1-CNN model of (a)HIF7S and HIF7SPTM and (b) HIF9S and HIF9SPTM. The gray curves overlapped on the feature weight histogram are the normalized POF of corresponding peptides. Yellow-shaded bands indicate spectra change peaks reflect the hydroxylation-induced spectra change and can be assigned to specific vibration modes, the red-shaded bands highlight characteristic citrate peaks, whereas gray-shaded bands indicate other spectral regions discussed in the main text.*

Gradient-weighted Class Activation Mapping(Grad-CAM) feature visualization has recently emerged as a useful tool for extracting spectral features from CNN-based spectroscopic models.[52, 53] In this work, we extended its application to single-molecule SERS spectral analysis. Normalized POF profiles were used as representative class-level inputs to explore whether recurrent spectral regions overlapped with CNN-sensitive regions. Figure 4 compares the feature weights extracted from the CNN model with the normalized POF profiles. Since POF profiles are transformed statistical representations rather than individual SERS spectra, the resulting Grad-CAM maps were interpreted qualitatively as class-level feature visualization.

The extracted feature-weight profiles exhibit groups of spikes, indicating Raman shift regions with high importance for discrimination. Notably, high-weight features with values ≥0.3 were observed at or near 680, 712, 1016, 1121, and 1234 $cm^{-1}$. These regions correspond well with the yellow-highlighted bands in Figure 2, where hydroxylation-induced spectral changes were associated with altered adsorption orientation. Additional high feature weights were observed around 994, 1073 $cm^{-1}$, and the amide II region, highlighted by gray shade bands. Although peak assignment in these regions is complicated by overlapping contributions from different residues, observable spectral changes suggest that these bands may still contribute to peptide hydroxylation discrimination.

It is worth noting that the 1073 $cm^{-1}$ band is characteristic of citrate. The high feature weight near this region suggests that citrate-related spectral changes may also contribute to the CNN-based discrimination. In this context, citrate should not simply be regarded as a spectral interference, but may participate cooperatively in the discrimination process, possibly through hydroxylation-dependent adsorption or surface-interaction effects. The role of citrate contribution will be further discussed in the discrimination of 15AA peptide hydroxylation.

Overall, the agreement between the POF profiles and CNN-derived feature weights indicates that these two analyses provide complementary information. The POF analysis identifies recurrent spectral features across single-molecule events, whereas Grad-CAM highlights spectral regions that are sensitive to the CNN classification decision. Their consistency supports the use of combined POF and CNN feature-weight for interpreting hydroxylation discrimination.

**Hydroxylation-Dependent Adsorption Behaviour in Longer 15AA Peptides**

We further tested a pair of 15AA peptides to assess the capability of our method for discriminating hydroxylation in longer peptide sequences. The normalized POF profiles of HIF15S and HIF15SPTM are shown in Figure 5(a). Both peptides exhibited peaks at 680 and 712 $cm^{-1}$, although the 712 $cm^{-1}$ band was less prominent than those observed for HIF7SPTM and HIF9SPTM. The presence of these peaks indicates successful adsorption of both peptides onto the AuNP surface. The coexistence of the 680 and 712 $cm^{-1}$ bands suggests the absence of a single dominant adsorption orientation, which may be attributed to the increased structural complexity and the greater number of possible binding sites in the 15AA peptides. For example, the two Met residues may serve as binding sites through sulfur–Au interactions. In addition, the aromatic rings of Tyr and Phe, as well as the peptide termini, may also act as possible anchoring points.

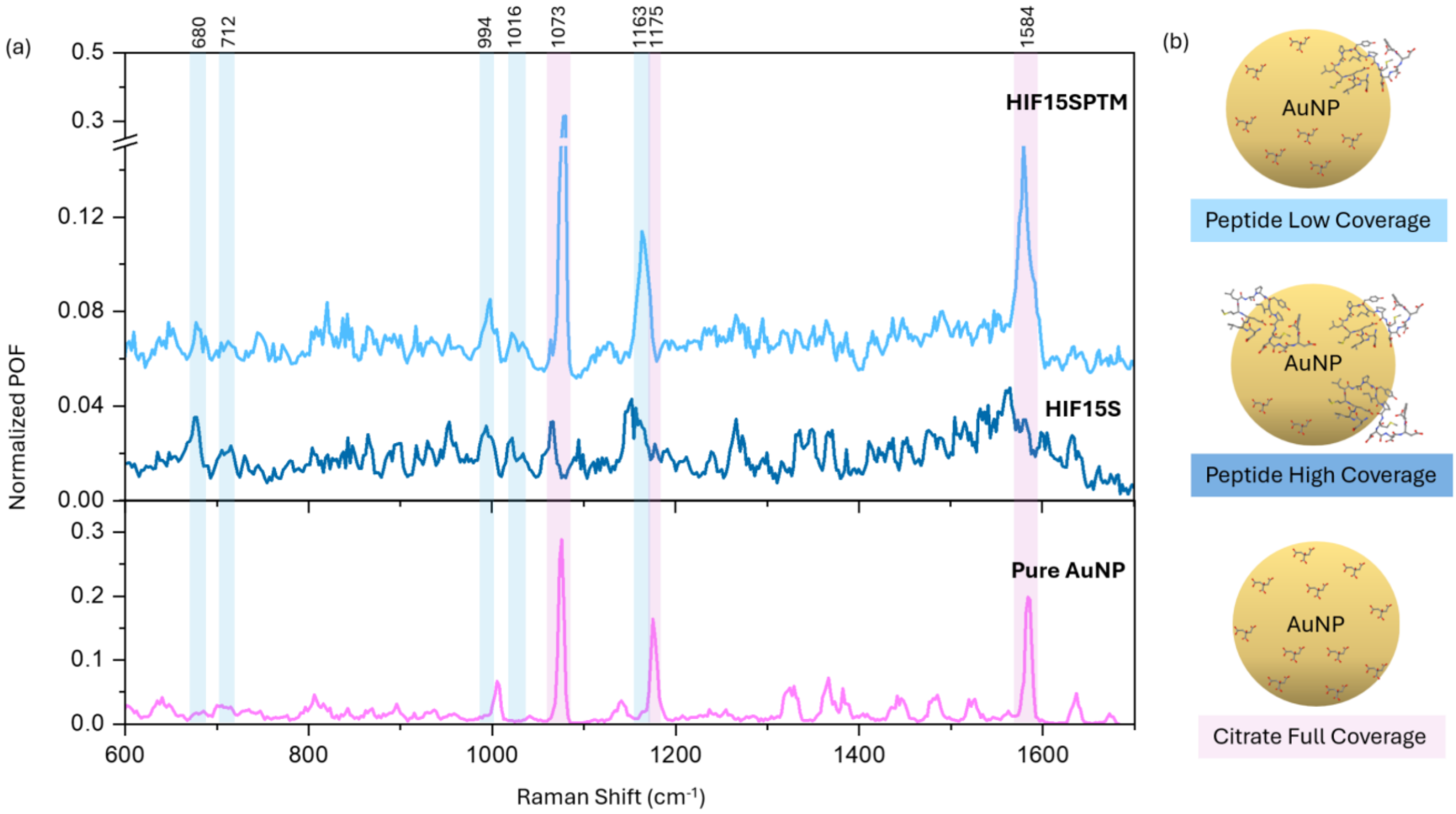

***Figure 5.*** *(a) Normalized POF profiles of HIF15S, HIF15SPTM, and pure AuNPs, shown in dark blue, light blue, and magenta, respectively. The blue-shaded bands indicate peptide-related peaks, while the red-shaded bands highlight characteristic citrate peaks. The offset of the HIF15SPTM POF curve is 0.05. (b) Proposed peptide adsorption states on citrate-covered AuNPs. The lower panel illustrates citrate-covered AuNPs without peptide adsorption, the middle panel represents relatively high surface coverage by HIF15S, and the upper panel represents relatively low surface coverage by HIF15SPTM.*

The peaks at 994, 1016, and 1163 $cm^{-1}$ further support peptide substitution of citrate on the AuNP surface, as these bands do not overlap with the POF peaks of citrate-capped AuNPs shown in the lower panel of Figure 5(a). Interestingly, the citrate-associated signature band at 1073 $cm^{-1}$ was much more pronounced for HIF15SPTM, with a normalized POF value of 0.266, approximately eight times higher than that observed for HIF15S at 0.033. Similarly, HIF15SPTM also showed a high POF near 1584 $cm^{-1}$, which overlaps with a citrate-related band. However, because this region is located within the amide II region and may overlap with peptide residue contributions, the 1073 $cm^{-1}$ band was selected as a more suitable indicator of citrate retention, as it is relatively less affected by peptide vibrational modes.

Although variations of peak height in 1073 $cm^{-1}$ nearby region were also observed in the 7AA and 9AA peptide pairs, the eight-fold difference between HIF15S and HIF15SPTM suggests a more pronounced hydroxylation-dependent difference in adsorption state for the 15AA peptides. The proposed adsorption states on citrate-covered AuNPs are illustrated in Figure 5(b). The lower panel represents citrate-covered AuNPs before peptide adsorption. The middle panel depicts higher surface coverage by HIF15S, while the upper panel illustrates lower surface coverage by HIF15SPTM. Under the lower-coverage condition of HIF15SPTM, a considerable amount of citrate may remain on the AuNP surface, resulting in the enhanced 1073 $cm^{-1}$ citrate band. The presence of hydroxylation may alter peptide conformation, charge distribution, or surface affinity, thereby reducing its adsorption efficiency on the AuNP surface. Among the three peptide pairs tested in this study, this effect appears most pronounced for the longer 15AA peptide, possibly because longer sequences can adopt more complex three-dimensional conformations.

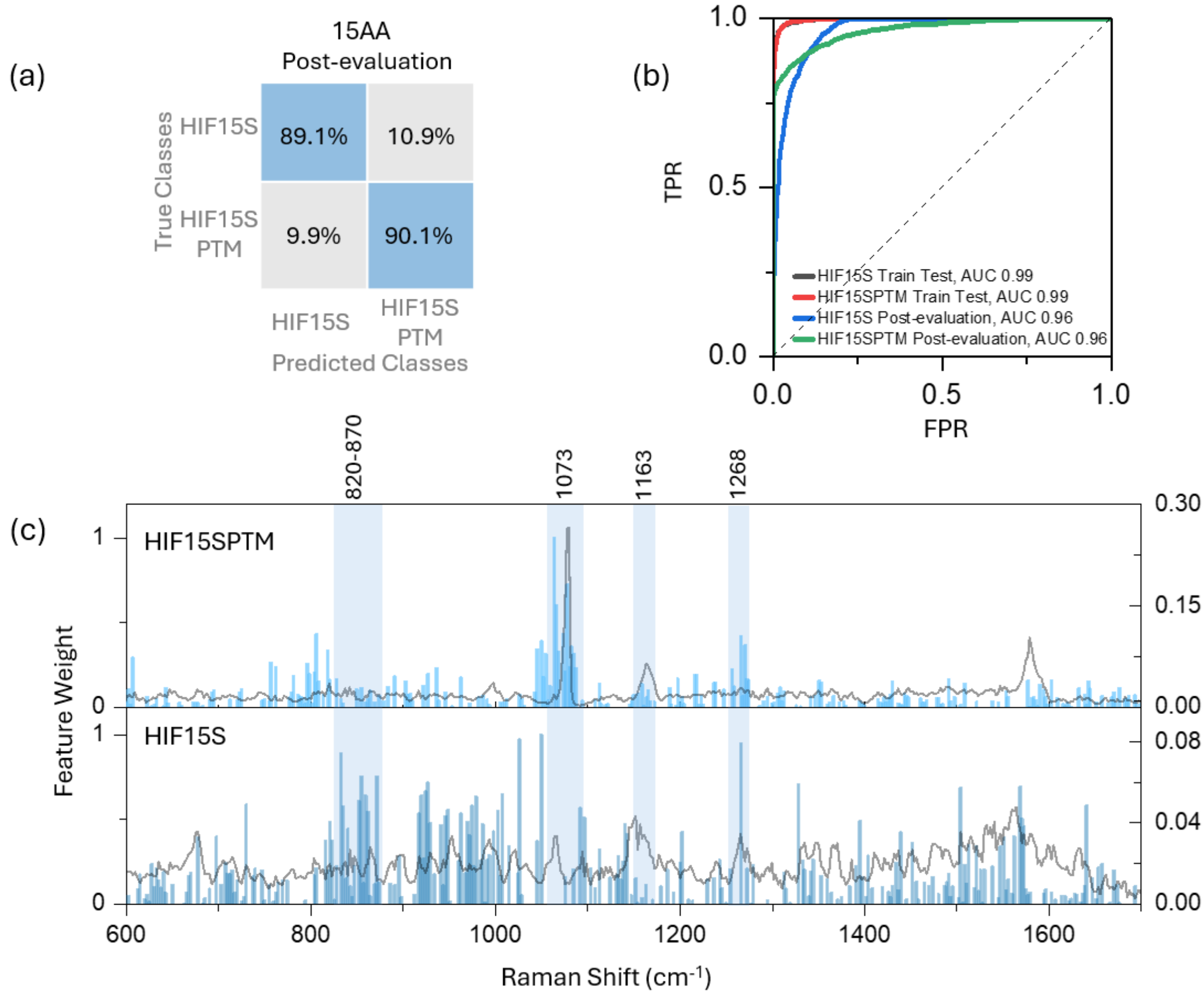

*__Figure 6.__ (a) Post-evaluation confusion matrix for HIF15S and HIF15SPTM, (b) ROC curves and AUC values for HIF15S and HIF15SPTM testing and post-evaluation datasets. (c) The feature weight extracted from the 1D-CNN model of HIF15S and HIF15SPTM. The gray-shaded bands highlight the spectra regions discussed in the main text.*

For CNN-based discrimination of the 15AA peptide pair, the model achieved an accuracy higher than 89% in both the training/testing process and post-evaluation. The corresponding post-evaluation confusion matrix is shown in Figure 6(a), while the ROC curve and AUC values are presented in Figure 6(b). The AUC value above 0.96 indicates reliable discrimination between HIF15S and HIF15SPTM. One might expect discrimination accuracy to decrease as peptide length increases, due to greater structural complexity and a reduced probability that the hydroxylation site directly enters the SERS hotspot which usually contains no more than 3 residues.[24] However, the results suggest that hydroxylation substantially affects the overall peptide property. In this case, the subtle molecular difference introduced by hydroxylation may be amplified by changes in local adsorption behaviour, thereby improving classification performance.

The feature-weight histogram is shown in Figure 6(c). A high feature weight near 1073 $cm^{-1}$ indicates that the hydroxylation-dependent change in adsorption behaviour contributes substantially to the CNN classification decision. In addition, high feature weights were observed in the 820-870, 1163, and 1268 $cm^{-1}$ regions. These bands do not overlap with the major citrate peaks and therefore reflect contributions from peptide structural or conformational changes. Together, the POF analysis, CNN classification results, and feature-weight visualization suggest that hydroxylation discrimination in the 15AA peptide is driven by both altered adsorption behaviour and intrinsic peptide spectral differences. Overall, these findings suggest that the combination of single-molecule SERS, POF analysis, and CNN-based feature visualization provides a promising strategy for discriminating subtle post-translational modifications in increasingly complex peptide systems.

**Conclusion**

In this work, we demonstrated single-molecule SERS discrimination of proline hydroxylation in native HIF-1α peptide fragments using a particle-in-pore plasmonic sensing platform combined with POF analysis, 1D-CNN classification. The particle-in-pore system enabled single-molecule SERS readout by trapping peptide-loaded AuNPs inside a gold nanopore and positioning the analyte within a highly confined plasmonic hotspot. Although hydroxylation introduces only a minor chemical modification, clear and reproducible differences were observed in the POF spectra of hydroxylated and non-hydroxylated peptides. For the 7AA and 9AA peptide pairs, hydroxylation suppressed the Met-associated band near 680 $cm^{-1}$ and promoted the appearance of bands near 712, 1016, 1121, and 1234 $cm^{-1}$, suggesting a hydroxylation-induced change from Met-dominated adsorption toward alternative adsorption configurations possibly involving Tyr or peptide terminal groups. CNN-based classification further confirmed that the single-molecule SERS spectra contained discriminative information related to peptide hydroxylation. Despite the intrinsic spectral heterogeneity of single-molecule SERS, the model achieved post-evaluation single-spectrum accuracies of 72.98%, 78.55%, and 89.74% for the 7AA, 9AA, and 15AA peptide pairs, respectively. The corresponding AUC values are 0.80, 0.88 and 0.96, respectively. Particularly, the value above 0.96 for the 15AA peptide pair indicates reliable discrimination. Feature-weight visualization showed that CNN-sensitive regions overlapped with POF-identified spectral changes, supporting the complementary use of statistical spectral reconstruction and deep learning interpretation. Noticeably, for the longer 15AA peptide pair, hydroxylation produced a more pronounced change in peptide adsorption behaviour, reflected by the substantially high citrate-associated band near 1073 $cm^{-1}$ in HIF15SPTM compared with HIF15S. This observation suggests that hydroxylation may alter peptide conformation, charge distribution, or surface affinity, thereby influencing citrate displacement and peptide surface coverage on AuNPs. Therefore, the discrimination mechanism may arise both from intrinsic peptide vibrational differences and hydroxylation-induced changes in adsorption state and local surface interactions. Overall, this study shows that the combination of particle-in-pore single-molecule SERS, POF profile analysis and CNN classification not only provides a sensitive strategy for detecting subtle peptide PTMs, but also demonstrates the potential of probing the biomolecule structure change, molecule-particle interaction, and local chemical environment change at the single-molecule level.

## Methods

### Materials

Gold nanoparticles (Sigma-Aldrich, 753645-25ML, $3.5 \times 10^{10}$ particles $mL^{-1}$) were used as received. The 7AA, 9AA and 15AA peptides, with and without proline hydroxylation, were purchased from Biomatik. According to the vendor, the peptide solubility was 0.5 mg $mL^{-1}$. Silicon wafers containing 100 nm thick SiN membranes were purchased from MicroChemicals GmbH. SYLGARD™ 184 Silicone Elastomer Kit was used for polydimethylsiloxane (PDMS) microfluidic channel casting and chip encapsulation.

### Nanopore Device Fabrication

The nanopore devices were fabricated following our previously reported procedure.[28] Briefly, gold nanoholes were fabricated on low-stress SiN membranes supported by silicon chips using focused ion beam (FIB) milling with an FEI Helios DualBeam system. The SiN window was $1 \times 1$ $mm^2$ in size and 100 nm thick, and the supporting silicon chips were $1 \times 1$ $cm^2$. A 2 nm titanium adhesion layer and a 100 nm gold layer were sputtered onto the front side of the SiN membrane, while a 20 nm gold layer was deposited on the back side. Nanopores with an average diameter of approximately 250 nm were then milled from the back side of the membrane by FIB. Scanning electron microscopy (SEM) was used to characterize the nanopore size and morphology from the front side. The nanopores were arranged in a $3 \times 3$ array to facilitate locating the pore position under the microscope during Raman measurements. Subsequently, conformal 5 nm $Al_2O_3$ coating was deposited on chips by atomic layer deposition in the Beneq TFS 200 at 110 ºC using $Al(CH_3)_3$ and $H_2O$ as precursors. Finally, the nanopore chips were embedded in a homemade PDMS microfluidic chamber.

### Peptide Attachment to AuNPs

All peptides used in the measurements were physically adsorbed onto AuNPs following protocols reported in our previous work.[22, 23] In the final solution used for Raman measurements, the AuNP concentration was $1.3 \times 10^{10}$ particles $mL^{-1}$, and the buffer concentration was 5% PBS at pH 7.4. Peptide stock solutions were mixed with AuNPs and PBS buffer, and the mixtures were stored at 4 °C for 48 h before Raman measurements to allow sufficient peptide adsorption onto the AuNP surface.

The final peptide concentrations were calculated according to previous literature[54] to obtain approximately monolayer coverage on each AuNP. This condition was chosen so that, in the particle-in-pore system, only one molecule is expected to occupy the plasmonic hotspot and generate a single-molecule SERS signal. Details of the concentration calculation are provided in Supplementary Note 2.

**Raman Measurement**

Raman measurements were performed using a Renishaw Raman imaging microscope equipped with a 785 nm laser and a 60× water-immersion objective. Time-series Raman spectra were collected with an accumulation time of 0.1 s per spectrum and a laser power of 47 mW. Before measurement, the nanopore devices were treated with $O_2$ plasma to render the chip surface and nanopore sidewalls hydrophilic. Peptide-incubated AuNP solutions were then introduced into the bottom chamber of the nanopore device, while 5% PBS buffer at pH 7.4 was added to the top chamber. Each peptide sample was measured using a separate nanopore device to avoid cross-contamination. During Raman acquisition, a transmembrane voltage between −1 and 1 V was applied using platinum electrodes to facilitate AuNP trapping in the nanopore. Raman signals were collected from at least 10 nanopores on each device.

**Raman Data Processing**

Raman data processing was performed using MATLAB R2022a. All raw spectra were first subjected to AirPLS baseline correction to flatten the spectral background. To define the noise threshold, 200 spectra were collected from blank nanopores without AuNP trapping. The average intensity of these blank spectra plus five times the standard deviation at each Raman shift was used as the final baseline threshold for identifying effective spectra. A spectrum was considered to contain a significant Raman signal only when its intensity exceeded this threshold. A second filtering criterion was applied to remove spectra with abnormally high intensities, which were typically attributed to large particle aggregates. The median intensity in the 300–600 $cm^{-1}$ region, where interference from amino acid bands is relatively limited, was used as an intensity indicator. An empirical threshold value of 500 was applied to exclude spectra with excessively high intensities. After removal of spectra with either too low or too high intensity, spectra containing prominent Raman peaks were further selected using the MATLAB findpeaks function. The selected spectra were then subjected to cosmic ray removal and normalization to generate the final effective spectra for subsequent analysis.

**POF Analysis**

To calculate the normalized POF profiles, the effective spectra were first smoothed to improve peak-finding accuracy. Peak positions were then identified using the MATLAB findpeaks function. The minimum peak prominence and minimum peak height were set to 0.1 and 0.35, respectively. The number of effective spectra used for POF analysis is listed in Table S3 in the Supporting Information. The normalized POF at each Raman shift was calculated by counting the number of spectra containing a peak at that Raman shift and dividing this value by the total number of effective spectra.

**CNN Model and Feature Visualization**

The detailed codes are available on GitHub:
https://github.com/zyqjulia-sketch/single-molecule-SERS-peptide-PTM-analysis/releases/tag/V0.1

Each effective SERS spectrum contained 981 spectral features and was used as the input to the model. The datasets were divided into training/testing and independent post-evaluation sets, as described in the main text. The numbers of effective spectra used for training, testing, and post-evaluation are summarized in Table S3. The classification accuracy and AUC values for the training/testing and post-evaluation datasets are listed in Table S4 in the Supporting Information. To improve model interpretability, one-dimensional gradient-weighted feature visualization was applied to identify spectral regions that contributed strongly to the CNN classification decision. The trained CNN model was first loaded, and the network architecture was modified to enable gradient calculation. Input spectral features were extracted and normalized using the pre-computed normalization parameters from the training workflow. The gradients of the classification loss with respect to the input features were then calculated using normalized POF profiles as a representative spectrum The loss function was defined as the categorical cross-entropy between the predicted and true class labels. The resulting normalized feature gradients were used to evaluate the relative contribution of each Raman shift to the CNN prediction, thereby providing insight into model-sensitive spectral regions.

## ASSOCIATED CONTENT

### Supporting Information

The Supporting Information is available free of charge on the ACS Publications website.

## AUTHOR INFORMATION

### Corresponding Author

Jianan Huang, Email: jianan.huang@oulu.fi

### Author Contributions

Yingqi Zhao fabricated the particle-in-pore devices, established the Raman measurement protocol, collected and analyzed the Raman spectra, developed scripts for Raman data pre-processing, and drafted the manuscript. Kuo Zhan developed scripts for Raman data pre-processing and effective spectrum selection, and established the 1D-CNN with Grad-CAM model for PTM discrimination. Yuge Liang assisted with SiN membrane fabrication and ALD deposition. Pei-Lin Xin contributed to the development of Raman spectra pre-processing scripts and performed DLS measurements. Enock Adjei Agyekum assisted with the analysis of the CNN model performance. Matti Putkonen assisted with ALD depositions. Francesco De Angelis contributed to device fabrication and revised the manuscript. Jianan Huang conceived the idea, acquired funding, supervised the work, and revised the manuscript. The manuscript was written with contributions from all authors, and all authors have approved the final version.

## ACKNOWLEDGMENT

This research receives support from Academy Research Fellow projects: TwoPoreProSeq (project number 347652), Biocenter Oulu spearhead project (LiveProSeq), MSCA CO-FUND Data4Healthcare program and the DigiHealth project (project number 326291), a strategic profiling project at the University of Oulu that is supported by the Academy of Finland and the University of Oulu. ALD Center Finland is acknowledged for providing the research infrastructure for ALD depositions. Jane and Aatos Erkko Foundation (Project: Novel materials for energy efficient microelectronics) is gratefully acknowledged for research funding.

## REFERENCES

1. Iommarini, L., et al., *Non-Canonical Mechanisms Regulating Hypoxia-Inducible Factor 1 Alpha in Cancer.* Frontiers in Oncology, 2017. **7**.

2. Ma, H., X.X. Han, and B. Zhao, *Enhanced Raman spectroscopic analysis of protein post-translational modifications.* TrAC Trends in Analytical Chemistry, 2020. **131**: p. 116019.
3. Semenza, G.L., *Hypoxia-inducible factors: mediators of cancer progression and targets for cancer therapy.* Trends in Pharmacological Sciences, 2012. **33**(4): p. 207-214.
4. Ponomarenko, E., et al., *The Size of the Human Proteome: The Width and Depth.* International Journal of Analytical Chemistry, 2016, 7436849, 6 pages, 2016**.**
5. Li, Z., et al., *Nanopore-Based High-Resolution Detection of Multiple Post-Translational Modifications in Protein.* Angewandte Chemie-International Edition, 2025. **64**(15).
6. Rosen, C., D. Rodriguez-Larrea, and H. Bayley, *Single-molecule site-specific detection of protein phosphorylation with a nanopore.* Nature Biotechnology, 2014. **32**(2): p. 179-181.
7. Ensslen, T., et al., *Resolving Isomeric Posttranslational Modifications Using a Biological Nanopore as a Sensor of Molecular Shape.* J Am Chem Soc, 2022. **144**(35): p. 16060-16068.
8. Zhang, L., et al., *Detection of single peptide with only one amino acid modification via electronic fingerprinting using reengineered durable channel of Phi29 DNA packaging motor.* Biomaterials, 2021. **276**: p. 121022.
9. Fahie, M.A. and M. Chen, *Electrostatic Interactions between OmpG Nanopore and Analyte Protein Surface Can Distinguish between Glycosylated Isoforms.* J Phys Chem B, 2015. **119**(32): p. 10198-206.
10. Chan Cao , et al., *Deep learning-assisted single-molecule detection of protein post-translational modifications with a biological nanopore*. ACS Nano 2024, 18, 2, 1504–1515.
11. Chang, J., et al., *Precise Identification of Native Peptides with Posttranslational Proline Hydroxylation by Nanopore.* Angewandte Chemie-International Edition, 2025. **64**(9).
12. Sundararajan, N., et al., *Ultrasensitive detection and characterization of posttranslational modifications using surface-enhanced Raman spectroscopy.* Analytical Chemistry, 2006. **78**(11): p. 3543-3550.
13. Iarossi, M., et al., *High-Density Plasmonic Nanopores for DNA Sensing at Ultra-Low Concentrations by Plasmon-Enhanced Raman Spectroscopy.* Advanced Functional Materials, 2023. **33**(41).
14. Khabarov, K., et al., *Raman identification of single nucleotides flowing through permeable plasmonic films.* Nature Communications, 2025. **16**(1): p. 9113.
15. Du, J., et al., *The role of phosphorylation and dephosphorylation of shell matrix proteins in shell formation: an in vivo and in vitro study.* Crystengcomm, 2018. **20**(27): p. 3905-3916.
16. Geddis, A., et al., *Label-Free SERS Sensors for Real-Time Monitoring of Tyrosine Phosphorylation*. Analytical Chemistry, 2024, 96, 45, 17978–17983.
17. Li, Y., et al., *Label-free in situ monitoring of protein tyrosine nitration in blood by surface-enhanced Raman spectroscopy.* Biosensors & Bioelectronics, 2015. **69**: p. 1-7.
18. Sjöberg, B., et al., *Methionine oxidation by hydrogen peroxide in peptides and proteins: A theoretical and Raman spectroscopy study.* Journal of Photochemistry and Photobiology B-Biology, 2018. **188**: p. 95-99.
19. Guerrero, A. and R. Aroca, *Surface-enhanced Raman scattering of hydroxyproline.* Journal of Raman Spectroscopy, 2012. **43**(4): p. 478-481.
20. Cárcamo, J., et al., *Proline and hydroxyproline deposited on silver nanoparticles. A Raman, SERS and theoretical study.* Journal of Raman Spectroscopy, 2012. **43**(6): p. 750-755.
21. Lee, H., et al., *Designing surface-enhanced Raman scattering (SERS) platforms beyond hotspot engineering: emerging opportunities in analyte manipulations and hybrid materials.* Chemical Society Reviews, 2019. **48**(3): p. 731-756.
22. Huang, J.-A., et al., *Multiplexed Discrimination of Single Amino Acid Residues in Polypeptides in a Single SERS Hot Spot.* Angewandte Chemie International Edition, 2020. **59**(28): p. 11423-11431.
23. Huang, J.-A., et al., *SERS discrimination of single DNA bases in single oligonucleotides by electro-plasmonic trapping.* Nature Communications, 2019. **10**(1): p. 5321.

24. Zhao, Y., et al., *Label-Free Optical Analysis of Biomolecules in Solid-State Nanopores: Toward Single-Molecule Protein Sequencing*. Acs Photonics, 2022. **9**(3): p. 730-742.
25. Lim, S., et al., *Unlocking the Full Potential of SERS: Merging Direct and Indirect Approaches for Enhanced Analysis of Multiplex Plasticizer Analogs in Matrices.* Angewandte Chemie-International Edition, 2025. **64**(40).
26. Tang, W., et al., *Digital Decoding of Multicomponent Protein Systems via Nanocavity-Confined Single-Molecule Raman Fingerprinting*. Angewandte Chemie-International Edition, 2026. **65**(6).
27. Khabarov, K., et al., *Single-Molecule Peptide Discrimination via Flow-Through SERS and Machine Learning*. Acs Photonics, 2026. **13**(7): p. 1840-1845.
28. Zhao, Y., et al., *Single-Molecule SERS Discrimination of Proline from Hydroxyproline Assisted by a Deep Learning Model.* Nano Letters, 2025, 25, 18, 7499–7506.
29. Abramson, J., et al., *Accurate structure prediction of biomolecular interactions with AlphaFold 3.* Nature, 2024. **630**(8016): p. 493-500.
30. Tien, M., et al., *Maximum Allowed Solvent Accessibilites of Residues in Proteins.* Plos One, 2013. **8**(11).
31. Franco-Ulloa, S., et al., *Dispersion state phase diagram of citrate-coated metallic nanoparticles in saline solutions.* Nature Communications, 2020. **11**(1).
32. Buglak, A. and A. Kononov, *Comparative study of gold and silver interactions with amino acids and nucleobases.* Rsc Advances, 2020. **10**(56): p. 34149-34160.
33. Zrimsek, A., et al., *Single-Molecule Chemistry with Surface- and Tip-Enhanced Raman Spectroscopy.* Chemical Reviews, 2017. **117**(11): p. 7583-7613.
34. Shin, H., et al., *Frequency-Domain Proof of the Existence of Atomic-Scale SERS Hot-Spots.* Nano Letters, 2018. **18**(1): p. 262-271.
35. Li, W., et al., *High-Throughput Single-Molecule Surface-Enhanced Raman Spectroscopic Profiling of Single-Amino Acid Substitutions in Peptides by a Gold Plasmonic Nanopore.* Acs Nano, 2024. **18**(29): p. 19200-19207.
36. Graff, M. and J. Bukowska, *Surface-enhanced Raman scattering (SERS) spectroscopy of enantiomeric and racemic methionine on a silver electrode-evidence for chiral discrimination in interactions between adsorbed molecules.* Chemical Physics Letters, 2011. **509**(1-3): p. 58-61.
37. Zhu, G., et al., *Raman spectra of amino acids and their aqueous solutions.* Spectrochimica Acta Part a-Molecular and Biomolecular Spectroscopy, 2011. **78**(3): p. 1187-1195.
38. Hernández, B., et al., *The relationship between the tyrosine residue 850-830 $cm^{-1}$Raman doublet intensity ratio and the aromatic side chain $x_1$ torsion angle.* Spectrochimica Acta Part a-Molecular and Biomolecular Spectroscopy, 2024, **308**, 123681.
39. Wolpert, M. and P. Hellwig, *Infrared spectra and molar absorption coefficients of the 20 alpha amino acids in aqueous solutions in the spectral range from 1800 to 500 $cm^{-1}$*. Spectrochimica Acta Part a-Molecular and Biomolecular Spectroscopy, 2006. **64**(4): p. 987-1001.
40. Smirnov, M., et al., *Interaction of gold and platinum nanoparticles with L-tyrosine in aqueous solution: Conformational and dynamic changes.* Journal of Molecular Liquids, 2025. **420**: p. 126822.
41. Stewart, S. and P. Fredericks, *Surface-enhanced Raman spectroscopy of amino acids adsorbed on an electrochemically prepared silver surface.* Spectrochimica Acta Part a-Molecular and Biomolecular Spectroscopy, 1999. **55**(7-8): p. 1641-1660.
42. Hernández, B., et al., *Vibrational Analysis of Amino Acids and Short Peptides in Hydrated Media. IV. Amino Acids with Hydrophobic Side Chains: L-Alanine, L-Valine, and L-Isoleucine.* Journal of Physical Chemistry B, 2009. **113**(10): p. 3169-3178.
43. Stewart, S. and P. Fredericks, *Surface-enhanced Raman spectroscopy of peptides and proteins adsorbed on an electrochemically prepared silver surface.* Spectrochimica Acta Part a-Molecular and Biomolecular Spectroscopy, 1999. **55**(7-8): p. 1615-1640.

44. Herlinger, A.W. and T.V. Long, II, *Laser-Raman and infrared spectra of amino acids and their metal complexes. III. Proline and bisprolinato complexes.* Journal of the American Chemical Society, 1970. **92**(22): p. 6481-6486.
45. SIIMAN, O., et al., *SURFACE-ENHANCED RAMAN-SCATTERING BY CITRATE ON COLLOIDAL SILVER.* Journal of Physical Chemistry, 1983. **87**(6): p. 1014-1023.
46. MUNRO, C., et al., *CHARACTERIZATION OF THE SURFACE OF A CITRATE-REDUCED COLLOID OPTIMIZED FOR USE AS A SUBSTRATE FOR SURFACE-ENHANCED RESONANCE RAMAN-SCATTERING.* Langmuir, 1995. **11**(10): p. 3712-3720.
47. Kurouski, D., et al., *Amide I vibrational mode suppression in surface (SERS) and tip (TERS) enhanced Raman spectra of protein specimens.* Analyst, 2013. **138**(6): p. 1665-1673.
48. Deckert-Gaudig, T., E. Rauls, and V. Deckert, *Aromatic Amino Acid Monolayers Sandwiched between Gold and Silver: A Combined Tip-Enhanced Raman and Theoretical Approach.* Journal of Physical Chemistry C, 2010. **114**(16): p. 7412-7420.
49. Hernández, B., et al., *Vibrational Analysis of Amino Acids and Short Peptides in Hydrated Media. VIII. Amino Acids with Aromatic Side Chains: L-Phenylalanine, L-Tyrosine, and L-Tryptophan.* Journal of Physical Chemistry B, 2010. **114**(46): p. 15319-15330.
50. Li, J., et al., *Machine learning using convolutional neural networks for SERS analysis of biomarkers in medical diagnostics.* Journal of Raman Spectroscopy, 2022. **53**(12): p. 2044-2057.
51. Lu, P., et al., *CNN-assisted SERS enables ultra-sensitive and simultaneous detection of Scr and BUN for rapid kidney function assessment.* Analytical Methods, 2023. **15**(3): p. 322-332.
52. Shi, G., et al., *1D Gradient-Weighted Class Activation Mapping, Visualizing Decision Process of Convolutional Neural Network-Based Models in Spectroscopy Analysis.* Analytical Chemistry, 2023. **95**(26): p. 9959-9966.
53. Xiong, C., et al., *Rapid and precise detection of cancers via label-free SERS and deep learning.* Analytical and Bioanalytical Chemistry, 2023. **415**(17): p. 3449-3462.
54. Tien, M., et al., *Maximum Allowed Solvent Accessibilites of Residues in Proteins.* Plos One, 2013. **8**(11): p. e80635.

# Supporting Information

# Label-free SERS Discrimination of Native Proline Hydroxylation at Single-molecule peptide by Deep Learning-assisted plasmonic nanopore

Yingqi Zhao[1,3] Kuo Zhan [1,3], Pei-Lin Xin[1,3], Yuge Liang[1,3], Enock Adjei Agyekum[4], Matti Putkonen[5], Shuai Li[4], Francesco De Angelis[6] and Jianan Huang [1,2,3,*]

[1] Research Unit of Health Sciences and Technology, Faculty of Medicine, University of Oulu, Aapistie 5 A, 90220 Oulu, Finland.

[2] Research Unit of Disease Networks, Faculty of Biochemistry and Molecular Medicine, University of Oulu, Aapistie 5 A, 90220 Oulu, Finland.

[3] Biocenter Oulu, University of Oulu, Aapistie 5 A, 90220 Oulu, Finland.

[4] The Biomimetics and Intelligent Systems (BISG) research unit, Faculty of Information Technology and Electronic Engineering, University of Oulu, Oulu, Finland

[5] Department of Chemistry, University of Helsinki, P.O.Box 55 (Chemicum, A.I. Virtasen aukio 1)

[6] Istituto Italiano di Tecnologia, Via Morego 30, 16163, Genoa, Italy.

*Email: jianan.huang@oulu.fi

## Supplementary Note 1. AlphaFold peptide structure plot

The three-dimensional structure of the peptide was predicted using AlphaFold3 Server.[1] The peptide sequence was submitted in one-letter amino acid code from the N-terminus to the C-terminus. The prediction was run using default parameters unless otherwise stated. Prediction confidence was assessed using the per-residue predicted local distance difference test, pLDDT, ipTM and PAE values. The AlphaFold-predicted three-peptide model showed high per-residue confidence, as indicated by high pLDDT values( listed in Table S1). However, the PAE matrix was predominantly green, suggesting moderate uncertainty in the relative positioning of residues across the peptide. Therefore, the predicted structure was interpreted as a plausible local conformational model rather than a definitive single stable structure.

*Table S1. The pLDDT values of AlphaFold peptide structure prediction*

| Peptide Name | pLDDT |
|---|---|
| HIF7S | >90 |
| HIF9S | >70 |
| HIF15S | >70 |

## Supplementary Note 2. Materials, Fabrication and CNN model Related Information

*Table S2. The DLS measured particle Size, Dispersity Index, and Zeta potential of AuNP before and after incubation with peptides.*

| Peptide Name | Size(nm) | Dispersity Index | Zeta Potential(mV) | Peptide Final Concentration(nM) | pH |
|---|---|---|---|---|---|
| HIF7S | 74.26 | 0.23 | -41.6 | 16.63 | 7.4 |
| HIF7SPTM | 60.64 | 0.13 | -43.0 | 16.63 | 7.4 |
| HIF9S | 57.16 | 0.11 | -43.3 | 13.77 | 7.4 |
| HIF9SPTM | 60.82 | 0.15 | -42.1 | 13.77 | 7.4 |
| HIF15S | 58.71 | 0.12 | -46.4 | 7.41 | 7.4 |
| HIF15SPTM | 58.73 | 0.12 | -44.4 | 7.41 | 7.4 |
| Unmodified AuNP | 60.48 | 0.13 | -42.7 | NA | 7.4 |

*Table S3. The summary of effective spectra number used for POF profile calculation, training and post-evaluation.*

| | **HIF7S** | **HIF7SPTM** | **HIF9S** | **HIF9SPTM** | **HIF15S** | **HIF15SPTM** |
|---|---|---|---|---|---|---|
| Number of effective spectra for POF profile calculation | 16380 | 16380 | 16380 | 6831 | 10421 | 12908 |
| Number of effective spectra for Training | 16380 | 16380 | 16380 | 6831 | 10421 | 12908 |
| Number of effective spectra for post-evaluation | 8891 | 9375 | 5638 | 2283 | 3193 | 5670 |

*Table S4. The summary of single-spectrum classification accuracy and AUC for training and post-evaluation data sets.*

| | **HIF7S** | **HIF7SPTM** | **HIF9S** | **HIF9SPTM** | **HIF15S** | **HIF15SPTM** |
|---|---|---|---|---|---|---|
| Accuracy of training | 96.1% | 94.1% | 95.8% | 87.2% | 97.8% | 97.8% |
| Accuracy of training test | 97.3% | 91.8% | 94.4% | 83.6% | 97.2% | 97.6% |
| Accuracy of post- evaluation | 72.8% | 73.2% | 77.8% | 80.3% | 89.1% | 90.1% |
| AUC of the training test | 0.97 | 0.97 | 0.97 | 0.97 | 0.99 | 0.99 |
| AUC of post- evaluation | 0.80 | 0.80 | 0.88 | 0.88 | 0.96 | 0.96 |
| Overall post-evaluation accuracy | 72.98% | | 78.55% | | 89.74% | |

## Supplementary Note 3. Calculation of amino acids (AA) and peptide concentration

The concentration of peptides required to achieve monolayers on the gold nanoparticle (AuNP) surfaces was determined by empirical values of maximum solvent accessibilities[2] of residues in proteins found in the literature[3] . The surface area of a single ϕ50 nm AuNP is calculated as 7850 $nm^2$. The amount of peptide molecules required to form a monolayer on a single AuNP was calculated by surface area of an unfolded peptide by $A=1.45 \times MW + 21$, where MW is the molecular weight of the peptide(unit Dalton)[4] The Maximum accessible surface area, number of molecules per AuNP to form a monolayer and final peptide concentration are summarized in Table S The calculation was based on the peptides without hydroxylation. The same concentration was used to prepare a peptide solution with and without hydroxylation with the same length. The calculation here assumes the peptide form; the peptides may fold in the solution and bind to the gold surface via a specific functional group. Therefore, a monolayer capping on the peptides on the AuNP is not guaranteed.

*Table S5: Maximum accessible surface area of peptides, number of peptides on one AuNP, and peptide final concentration*

| Peptide Name | Maximum accessible surface area ( $Å^2$) | Number of peptides on one AuNP for monolayer coverage | Peptide Final Concentration(nM) |
|---|---|---|---|
| HIF7S | 12.10 | 617 | 16.63 |
| HIF7SPTM | 12.10 | 617 | 16.63 |
| HIF9S | 15.38 | 510 | 13.77 |
| HIF9SPTM | 15.38 | 510 | 13.77 |
| HIF15S | 26.09 | 275 | 7.41 |
| HIF15SPTM | 26.09 | 275 | 7.41 |

**Reference**

(1) Abramson, J.; Adler, J.; Dunger, J.; Evans, R.; Green, T.; Pritzel, A.; Ronneberger, O.; Willmore, L.; Ballard, A. J.; Bambrick, J.; et al. Accurate structure prediction of biomolecular interactions with AlphaFold 3. *Nature* **2024**, *630* (8016), 493-500.
(2) Tien, M.; Meyer, A.; Sydykova, D.; Spielman, S.; Wilke, C. Maximum Allowed Solvent Accessibilites of Residues in Proteins. *Plos One* **2013**, *8* (11), e80635, Article.
(3) Huang, J.-A.; Mousavi, M. Z.; Giovannini, G.; Zhao, Y.; Hubarevich, A.; Soler, M. A.; Rocchia, W.; Garoli, D.; De Angelis, F. Multiplexed Discrimination of Single Amino Acid Residues in Polypeptides in a Single SERS Hot Spot. *Angewandte Chemie International Edition* **2020**, *59* (28), 11423-11431.
(4) Miller, S.; Lesk, A. M.; Janin, J.; Chothia, C. The accessible surface area and stability of oligomeric proteins. *Nature* **1987**, *328* (6133), 834-836.